\documentclass[aps,prl,twocolumn,groupedaddress,showpacs]{revtex4-1}
\usepackage{color}
\usepackage{amsmath}
\usepackage{graphicx}
\usepackage{bm}
\usepackage[normalem]{ulem}

\newcommand\grad{{\boldsymbol{\nabla}}}
\newcommand\bN{{\mathbf{N}}}
\newcommand\bn{{\mathbf{n}}}
\newcommand\bu{{\mathbf{u}}}
\newcommand\bv{{\mathbf{v}}}
\newcommand\bp{{\mathbf{p}}}
\newcommand\bx{{\mathbf{x}}}
\newcommand\bP{{\mathbf{P}}}
\newcommand\btheta{{\bm{\theta}}}
\newcommand\tD{{\bm{\mathsf{D}}}}

\newcommand\tE{{\bm{\mathsf{E}}}}
\newcommand\tW{{\bm{\mathsf{W}}}}
\newcommand\tI{{\mathbf{{I}}}}
\newcommand\tR{{\bm{\mathsf{R}}}}
\newcommand\tQ{{\bm{\mathsf{Q}}}}

\newcommand\te{{\bm{\mathsf{e}}}}
\newcommand\tSigma{{\bm{\Sigma}}}

\newcommand\transpose{{\mathsf{T}}}
\newcommand\opL{{\mathcal{L}}}

\renewcommand{\eqref}[1]{Eq.~(\ref{#1})}

\begin{document}

%\preprint{}

\title{Spontaneous Circulation of Confined Active Suspensions}


\author{Francis G. Woodhouse}
\author{Raymond E. Goldstein}
%\email[For correspondence: ]{R.E.Goldstein@damtp.cam.ac.uk}
\affiliation{Department of Applied Mathematics and Theoretical Physics, Centre for 
Mathematical Sciences,\\ University of Cambridge, Wilberforce Road, Cambridge CB3 0WA, 
United Kingdom}

\date{\today}

\begin{abstract}
Many active fluid systems encountered in biology are set in total geometric confinement.  
Cytoplasmic streaming in plant cells is a prominent and ubiquitous example, in which cargo-carrying 
molecular motors move along polymer filaments and generate coherent cell-scale flow. When 
filaments are not fixed to the cell periphery, a situation found both \textit{in vivo} and 
\textit{in vitro}, we observe that the basic dynamics of streaming are closely related to those
of a non-motile stresslet suspension. Under this model, it is demonstrated that confinement makes 
possible a stable circulating state; a linear stability analysis reveals an activity threshold 
for spontaneous auto-circulation. Numerical analysis of the long-time behavior reveals a phenomenon akin to
defect separation in nematic liquid crystals, and a high-activity bifurcation to an oscillatory regime.
\end{abstract}

%\keywords{}

\maketitle

\begin{figure}[b]
 \includegraphics[width = 0.90\columnwidth]{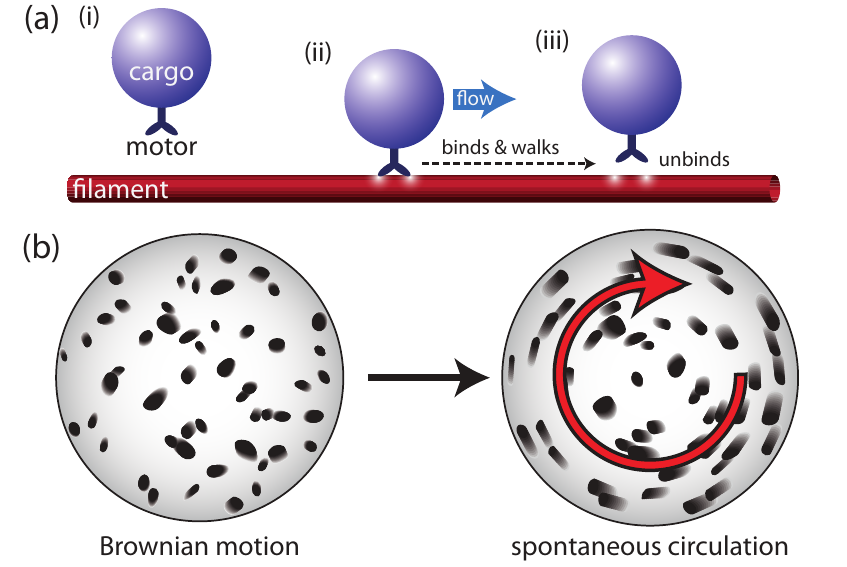}
\caption{(color online).  Cytoplasmic streaming {\it in vivo} and {\it ex vivo}.  (a)  A molecular motor attached to a 
vesicle (i) encounters a filament, (ii) binds and walks along it, entraining fluid, before 
(iii) unbinding stochastically. (b) A drop of cytoplasm extracted 
from a plant cell transitions from random Brownian fluctuations to ordered circulation 
\cite{Yotsuyanagi}.}
\label{fig:droplet_diagram}
\end{figure}

Cytoplasmic streaming is the deliberate, driven motion of the entire contents of large 
eukaryotic cells. It is effected by cargo-laden molecular motors walking along polymer filaments 
and entraining the surrounding fluid (Fig.~\ref{fig:droplet_diagram}a); the combined action of many 
of these motors can generate flow speeds in excess of $100$ $\mathrm{\mu m/s}$ for certain 
freshwater algae. While inroads are being made into understanding its function 
\cite{Protoplasma,Goldstein2008}, surprisingly little is known about how it is initially 
established within cells.

In a remarkable, yet apparently little-known investigation into the development of streaming, 
Yotsuyanagi \cite{Yotsuyanagi} in 1953 examined isolated droplets of 
cytoplasm forcibly extracted from algal cells. He observed a progression from isolated 
Brownian fluctuations to a coherent, global circulation of the entire droplet 
contents (Fig.~\ref{fig:droplet_diagram}b). However, we need not limit ourselves to 
\textit{ex vivo} experiments: Kamiya \cite{Kamiya1959} describes a similar blooming of rotational 
cyclosis in the development of \textit{Lilium} pollen cells, and Jarosch \cite{Jarosch1956} quantitatively 
analyzed the same disorder-to-order transition occurring within \textit{Allium} cells over the course 
of a few hours. Based on these observations, one is led to ask: is it possible that a simple 
self-organization process could lie at the heart of streaming?

When the filaments are not locked in position, as is likely in Yotsuyanagi's experiments, a 
cargo-carrying motor walking on a free filament constitutes a force dipole. Therefore, 
these cytoplasmic dynamics belong to the burgeoning field of \emph{active fluids}.  
With roots in self-organizing flocking models \cite{Flocking}, an active fluid is a suspension of 
force dipoles interacting via short- and long-range forces: a 
system like a liquid crystal but with continuous injection of energy at the microscale. Such systems generically 
possess spontaneous flow instabilities \cite{SimhaRamaswamy2002} and can exhibit complex patterns and 
flows \cite{Voituriez2005}, 
including asters and vortices
\cite{Kruse2004, Elgeti2011, Kruse2005}, laning \cite{GiomiMarchetti2011, Giomi} and density 
waves \cite{Kruse2005, GiomiMarchetti2011, Giomi}.
Spontaneous flow in particular is a key characteristic of flocking dynamics \cite{Flocking}, and 
assumes an important role in applications such as cortex remodeling processes \cite{Salbreux2008}.

Despite the ubiquity of relevant situations, of which streaming is a major example, the influence 
of total confinement is relatively little-studied. Kruse {\it et al.} \cite{Kruse2004} included a finite domain 
in their study of single defect stability in polar gels, assuming perfect alignment everywhere bar the defect 
core,  and found a dependence of aster and vortex stability on domain size.
More recently, Schaller {\it et al.} \cite{Schaller2010} underlined the critical importance of long-range 
hydrodynamics in confined 
systems: swirling patterns were observed experimentally in a totally confined actin motility assay, 
but were absent in cellular automaton 
simulations. They concluded that confined flows are responsible for the formation 
and stability of the global circulation. It seems reasonable, therefore, to posit that combining confinement 
with the spontaneous flow characteristics of active fluids will lead to circulatory streaming states. Indeed, 
such effects were taken advantage of by F{\"u}rthauer {\it et al.} \cite{Furthauer2012} to construct 
theoretically a `Taylor--Couette motor', albeit in a geometry topologically distinct from a single confined chamber.

Through theory and simulation, we show here 
that the combination of confinement and activity allows for the emergence of stable 
self-organized rotational streaming.
This is achieved using a ground-up approach employing closure techniques new to active suspension theory.
Our model assumes that 
short, rigid filaments are suspended in a Newtonian, zero Reynolds number fluid, 
and they exert extensile, or `pusher', dipolar forces
on the fluid; this can be viewed as the effect of processive molecular
motors landing randomly along a filament and walking toward one end,
implying an average motor location forward of the filament midpoint.
The suspension is taken to be dilute, so filaments interact via hydrodynamics 
only, and is confined within a no-slip sphere of diameter $L$.

Working in $d$ dimensions we generalize the standard kinetic approach to these systems 
\cite{SaintillanShelley}. 
The spatial and angular distribution function $\Psi(\bx,\bp,t)$ of the filaments, where $|\bp|=1$,
satisfies a Smoluchowski equation
\begin{align}\label{eq:smol}
\frac{\partial \Psi}{\partial t} = -\grad_\bx \cdot (\dot\bx \Psi) - \grad_\bp \cdot (\dot\bp \Psi)
\end{align}
where $\grad_\bx \equiv \partial/\partial\bx$ and $\grad_\bp \equiv (\tI - \bp\bp)\cdot\partial/\partial\bp$. 
The spatial and rotational fluxes are
\begin{align*}
\dot\bx &\equiv \bu + V\bp - \tD^{(s)} \cdot \grad_\bx \log \Psi, \\
 \dot\bp &\equiv (\tI - \bp \bp) \cdot (\gamma \tE + \tW) \cdot \bp - D^{(r)} \grad_\bp \log \Psi,
\end{align*}
where $V$ is a self-advection speed, $\gamma \in [-1,1]$ is a shape parameter ($\gamma \rightarrow 1$ 
for a slender rod), $\tD^{(s)}$ is a spatial diffusion tensor and $D^{(r)}$ is a rotational diffusion constant. 
The fluid has velocity field $\bu$, rate-of-strain tensor $\tE \equiv (\grad\bu + \grad\bu^\transpose)/2$ 
and vorticity tensor $\tW \equiv (\grad\bu - \grad\bu^\transpose)/2$. The filament pusher stresslet 
of strength $\sigma > 0$ generates a stress tensor 
$\tSigma \equiv -\sigma \int_\bp d\bp \, (\bp\bp - \tI/d)\Psi$ that drives fluid flow by the 
Stokes equation $-\mu\nabla^2 \bu + \grad \Pi = \grad \cdot \tSigma$ with viscosity $\mu$ and pressure 
$\Pi$, subject to incompressibility $\quad \grad \cdot \bu = 0$. Confinement induces the no-slip 
boundary condition $\bu=0$ on $|\bx| = L/2$.

While simulations of the full system (\ref{eq:smol}) are 
possible \cite{SaintillanShelley, HelzelOtto2006, PahlavanSaintillan2011}, 
here we develop evolution equations for the primary orientation moments 
\cite{DoiEdwards1988, DeGennesProst1995, BaskaranMarchetti2008}. Given the orientational average 
$\langle \phi \rangle \equiv \int_\bp d\bp \, \phi \Psi$, define the concentration $c \equiv \langle 1 \rangle$, 
polar moment $\bP \equiv \langle\bp\rangle$ and nematic moment 
$\tQ \equiv \langle \bp\bp - \tI/d \rangle$. Equations of motion for these fields in terms of higher moments
can then be derived by taking appropriate weighted integrals of 
\eqref{eq:smol} \cite{WoodhouseGoldsteinToAppear}.

We pare down complications by specializing to two dimensions ($d=2$), rodlike particles ($\gamma=1$) 
and isotropic diffusion ($\tD^{(s)} = D^{(s)} \tI$), and neglect self-advection ($V\equiv 0$). This 
last assumption decouples the $c$ dynamics into pure advection-diffusion and eliminates all polar 
interactions, so we take a constant concentration $c\equiv c_0$ and neglect $\bP$. However, the
 remaining $\tQ$ dynamics still depends on the fourth moment contraction $\langle\bp\bp\bp\bp\rangle : \tE$, 
and a closure is needed. Typically this is done by taking the distribution $\Psi$ to be a functional purely of 
the first three moments, yielding a closure linear in $\tQ$ \cite{BaskaranMarchetti2008}. In dense 
active systems this is permissible, owing to the presence of local interaction terms in higher powers 
of $\tQ$; here, however, it is 
the above fourth moment term which provides all stabilizing nonlinearities, so greater care must be taken. 
Instead we adopt a new approach by adapting
a closure of Hinch and Leal \cite{HinchLeal1976} to $d=2$, yielding
\begin{align*}
\langle\bp\bp\bp\bp\rangle : \tE
 \approx \frac{1}{4c_0} \left[ 4 \tQ \cdot \tE \cdot \tQ + 2c_0(\tE \cdot \tQ + \tQ \cdot \tE) \right. \\
\quad + \left. c_0^2 \tE - 2\tI \tQ^2 : \tE \right].
\end{align*}
This is derived in \cite{HinchLeal1976} as an interpolation between exact closures for the regimes of total 
order and disorder, giving a simple approximation to the hydrodynamic nonlinearities.
After non-dimensionalizing by rescaling $\bx \rightarrow L\bx$, $t \rightarrow (c_0L^2/\mu)t$, 
$\bu \rightarrow (\mu/c_0L)\bu$, $\Pi \rightarrow (\mu^2/c_0L^2)\Pi$, 
$\tSigma \rightarrow (c_0L^2/\mu^2)\tSigma$ and $\tQ \rightarrow c_0 \tQ$, the final model reads
\begin{align} \label{eq:model}
\frac{D\tQ}{Dt} = d^{(s)} \nabla^2 \tQ - 4 d^{(r)} \tQ + \tfrac{1}{2}\tE - 2\tQ \cdot \tE \cdot \tQ
\end{align}
where $D/Dt \equiv \partial/\partial t + \bu \cdot \grad$, with non-dimensional diffusion 
constants $d^{(s)} \equiv (c_0/\mu)D^{(s)}$ and $d^{(r)} \equiv (c_0 L^2 /\mu)D^{(r)}$. 
This is subject to the Stokes equation $-\nabla^2 \bu + \grad \Pi = -\sigma_0 \grad \cdot \tQ$ 
and incompressibility $\grad \cdot \bu = 0$ with non-dimensional dipole stress 
$\sigma_0 \equiv (c_0 L/\mu)^2 \sigma$. The fluid boundary condition reads 
$\bu = 0$ on $|\bx| = 1/2$.
Among the variety of admissible boundary conditions on $\tQ$ we focus here on the \emph{natural} 
condition $\bN \cdot \grad \tQ = 0$, where $\bN$ is the boundary normal vector. Qualitatively 
similar results are found with fixed boundary-parallel or 
boundary-perpendicular conditions \cite{WoodhouseGoldsteinToAppear}.

The model (\ref{eq:model}) has the structure of a Landau theory for the order
parameter $\tQ$. As $\tE$ is linear in the velocity $\bu$, and $\bu$ is (nonlocally) linear
in $\sigma_0\tQ$ via the Stokes equation, the term $(1/2)\tE\propto \sigma_0\tQ$.   
It follows in the usual manner that there is an effective linear
term in $\tQ$ that will become positive for sufficiently large activity $\sigma_0$ relative to 
$-4d^{(r)}$.  If this
is sufficient to overcome the diffusive stabilization $d^{(s)}$ then the
amplitude of the ensuing instability will be limited by the nonlinear term $2\tQ\cdot\tE\cdot\tQ\propto \tQ^3$.

We first seek a steady non-flowing axisymmetric state $\tQ^0$.  In polar coordinates 
$\alpha = (r,\theta)$ the tensor Laplacian of $\tQ$ has 
primary components
\begin{align*}
(\nabla^2 \tQ)_{r\alpha} = \opL Q_{r\alpha}
\equiv
\frac{1}{r} \frac{\partial}{\partial r} \left( r \frac{\partial Q_{r\alpha}}{\partial r} \right) - \frac{4}{r^2} Q_{r\alpha}~,
\end{align*}
while the others follow from symmetry and the tracelessness of $\tQ$. \eqref{eq:model} therefore implies 
$Q^0_{rr}$ and $Q^0_{r\theta}$ each satisfy a (modified) Bessel equation in $z \equiv 2\Delta r$, \textit{viz.} 
$z^2 \partial_z^2 Q^0_{r\alpha} + z \partial_z Q^0_{r\alpha} - (z^2 + 4)Q^0_{r\alpha} = 0$, where 
$\Delta^2 \equiv d^{(r)}/d^{(s)}$. Thus $Q^0_{r\alpha} \propto I_2(2\Delta r)$;
since $I_2$ is monotonic, the boundary conditions imply $\tQ^0 = 0$ everywhere.

\begin{figure}[t]
 \includegraphics[width = 0.95\columnwidth]{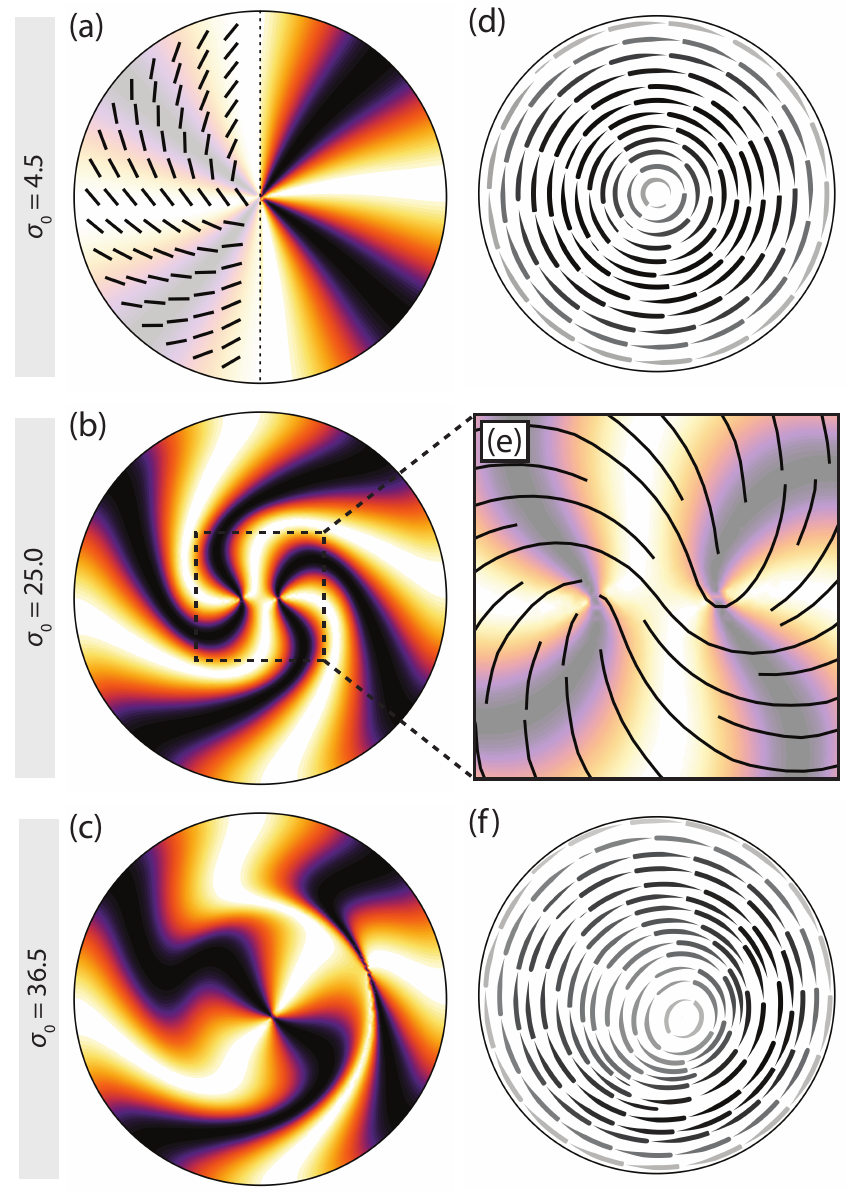}
\caption{(color online).  Numerical results beyond the spontaneous circulation threshold. (a,b,c) Simulated 
schlieren textures of nematic order director $\bn$ (i.e. density plot of $(n_x n_y)^2$). Lighter 
corresponds to diagonally-oriented filaments, darker to horizontal or vertical. (a) Steady circulation 
with a central spiral defect at low activity (left overlay shows order field $\bn$),
(b) steady central defect separation into a pair of hyperbolic 
defects, (c) snapshot of oscillatory behavior with widely separated mobile defects. (d) Flow streamlines 
for low activity, showing circulation about the system center; darker streamlines indicate faster 
flow. (e) Enlargement of nematic director field structure in texture (b), showing two hyperbolic defects. 
(f) Flow streamlines for high-activity oscillation associated with texture (c) exhibiting off-center flow circulation. 
In all cases, $d^{(r)} = d^{(s)} = 0.025$.}
\label{fig:behaviour_spectrum}
\end{figure}

Now, perturb axisymmetrically: let $\tQ = \epsilon \tR$, $\epsilon \ll 1$, and write $\bu = \epsilon v \hat\btheta$, 
$\tE = \epsilon \te$ for the induced flow (which has no radial component by incompressibility). Seek an 
exponentially growing state such that $\partial_t \tR  = s \tR$. Then to $O(\epsilon)$, the perturbation 
obeys $s \tR = d^{(s)}\nabla^2 \tR - 4d^{(r)} \tR + \tfrac{1}{2}\te$. To determine $\te$ we employ 
the technique of Kruse {\it et al.} \cite{Kruse2004} and write the Stokes equation as 
$\grad \cdot (-\tilde\Pi \tI + 2\te - \sigma_0 \tR) \equiv \grad \cdot \tSigma^\text{tot} = 0$. The 
$r$-component determines $\Pi$. The $\theta$-component reads 
$\partial_r \Sigma^\text{tot}_{r\theta} + (2/r)\Sigma^\text{tot}_{r\theta} = 0$, so for 
$\Sigma^\text{tot}_{r\theta}$ analytic at $r=0$ we find $\Sigma^\text{tot}_{r\theta} = 0$, i.e. 
$e_{r\theta} = (\sigma_0/2)R_{r\theta}$. Finally, $e_{rr} = 0$ as there is no radial velocity component. 
The perturbation therefore satisfies 
\begin{align}
d^{(s)} \opL R_{rr} &= (4d^{(r)} + s) R_{rr}, \label{eq:perturbation_rr} \\
d^{(s)} \opL R_{r\theta} &= \left(4d^{(r)} + s - \tfrac{\sigma_0}{4}\right) R_{r\theta}, \label{eq:perturbation_rt}
\end{align}
which are still of Bessel form. When $s > -4d^{(r)}$, \eqref{eq:perturbation_rr} has a solution in 
terms of $I_2$, so boundary conditions imply $R_{rr} = 0$. Now, let 
$\lambda \equiv (4d^{(r)} + s - \sigma_0/4)/d^{(s)}$ and write \eqref{eq:perturbation_rt} as 
$\opL R_{r\theta} = \lambda R_{r\theta}$. For $\lambda > 0$ this again gives solutions in terms 
of $I_2$ and so $R_{r\theta} = 0$. However, for $\lambda < 0$ (i.e. $\sigma_0$ sufficiently large) the solution
 is instead $R_{r\theta} \propto J_2(\sqrt{-\lambda} r)$. Applying the boundary condition $R_{r\theta}'(1/2) = 0$ 
yields the eigenvalue $\lambda = -4 y_0^2$ in terms of $y_0 \approx 3.054$, the first positive point 
satisfying $J_2'(y_0) = 0$. This implies that the homogeneous disordered state is unstable to
a spontaneously flowing mode when 
$\sigma>\sigma^*$,
%\footnote{Note that searching for solutions with higher modes $y_i > y_0$ 
%with $J'_2(y_i) = 0$ yields a greater condition on $\sigma_0$, so stronger activity begins to incorporate higher 
%modes, but the base mode dominates stability considerations}, 
where (in physical units)
\begin{align}\label{eq:natural_instability_condition}
\sigma^* \simeq \frac{16\mu}{c_0} \left( 9.33 \frac{D^{(s)}}{L^2} + D^{(r)}\right),
\end{align}
which we verified numerically by simulations of Eq. (2).

Stability analysis
in the unbounded case elicits instability of a long-wavelength band when $\sigma_0 > 16d^{(r)}$ (see also \cite{SaintillanShelley}), compatible with the
$L \rightarrow \infty$ limit of \eqref{eq:natural_instability_condition}. This illustrates the action 
of confinement as a strong constraint on the available excitation modes, allowing for selection of a 
single circulation mode as opposed to a band of wavenumbers. Similar spontaneous flows have 
been observed in active nematic models under periodic conditions \cite{Giomi} but the excited modes 
exhibit `laning' flows as opposed to the circulation seen here.
To lend perspective, we consider typical values of the material properties.  The stress
amplitude can be expressed as $\sigma=f\ell$, where $f$ is the (typically pN) force exerted by motors 
and $\ell$ is the (typically $\mu$m) separation of the opposing forces of the stresslet.
For micron-size rods we expect $D^{(r)}\sim 0.01$ s$^{-1}$ and 
$D^{(s)}\sim 10^{-9}$ cm$^{2}$/s, so for system sizes $L\gtrsim 10$ $\mu$m rotational diffusion
dominates in \eqref{eq:model}.  Then for a fluid of the viscosity of water 
in an idealized slab geometry
the instability will set in at $c_0\gtrsim 10^8$ cm$^{-3}$,
corresponding to a volume fraction well below $10^{-3}$.

\begin{figure}[t]
 \includegraphics[width = 0.95\columnwidth]{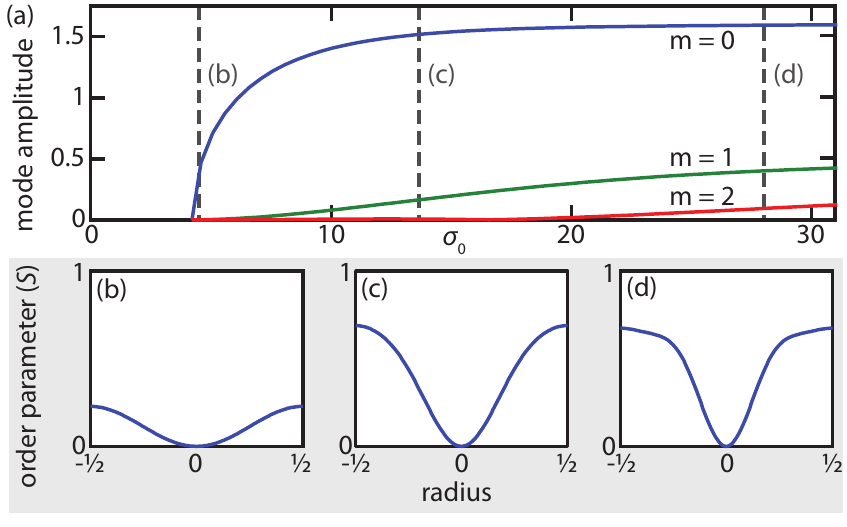}
\caption{(color online).  Details of the bifurcation to circulation.  (a) Numerically-evaluated steady state 
amplitudes $|S_0^{(m)}|$ of Bessel series expansion for 
axisymmetric part of order parameter $S$ at varying activity $\sigma_0$, with $d^{(r)} = d^{(s)} = 0.025$. 
(b-d) Profiles of $S$ at indicated points (b-d) in (a).}
\label{fig:axisymmetric_amplitudes}
\end{figure}

In order to confirm that the circulating configuration is steady at long times, we must turn to simulations 
of the fully nonlinear dynamics.  In the following numerical studies we vary the 
dipolar activity $\sigma_0$ while fixing the diffusion constants at 
$d^{(r)} = d^{(s)} = 0.025$, and use the eigendecomposition $\tQ = S(\bn\bn - \tI/2)$, where the 
order parameter $S$ and (headless) director $\bn$ are the degree of local alignment and the 
average alignment direction, respectively.
For sufficiently weak activity above $\sigma^*$, a stable steady state emerges of circulation about the system center 
(Fig.~\ref{fig:behaviour_spectrum}a\&d). The spiral pattern of the nematic director field is reminiscent of the 
predictions of Kruse {\it et al.} \cite{Kruse2004} for polar systems 
\citep[see also][]{Marenduzzo2010, Elgeti2011}.  Higher mode contributions can be examined by 
expanding the order parameter as $S(r,\theta) = \sum_n S_n(r) e^{in\theta}$, and applying 
an appropriate
$n=0$ mode expansion $S_0(r) = \sum_{m=0}^\infty S_0^{(m)} J_2(2 y_m r)$ where 
$J_2'(y_m) = 0$ and $y_m < y_{m+1}$. 
%\footnote{The functions $\{J_2(2 y_m r)\}_{m\geq 0}$ constitute a 
%complete set for functions $f(r)$ on $r \in [0,1/2]$ with $f'(1/2) = 0$; cf. a Bessel series for functions with 
%$f(1/2) = 0$.}. 
Mode amplitudes are extracted using orthogonality of the radial basis.
Figure~\ref{fig:axisymmetric_amplitudes} shows the steady-state values of the first three amplitudes 
$|S_0^{(m)}|$ as functions of $\sigma_0$. 

At larger values of $\sigma_0$, the steady state exhibits \emph{defect separation}: the central 
axisymmetric spiral defect in the nematic director field (with topological charge $+1$) splits into 
two closely spaced hyperbolic defects (each of charge $+1/2$), illustrated in 
Fig.~\ref{fig:behaviour_spectrum}b\&e. The system still possesses fluid 
circulation about the central axis, due to the symmetric positioning of the defects.  Defect separation
is perhaps unsurprising if we make contact with liquid crystal theory; for approximately isolated 
defects, the free energy penalty per defect is proportional to the square of its topological 
charge \cite{LubenskyProst1992}, rendering two $+1/2$ defects favorable over a single $+1$ spiral. 
Indeed, de las Heras {\it et al.} \cite{delasHeras2009} recently investigated the equivalent confined setup 
for a microscopic 
two-dimensional liquid crystal and always encountered defect separation.

\begin{figure}
\includegraphics[width=0.95\columnwidth]{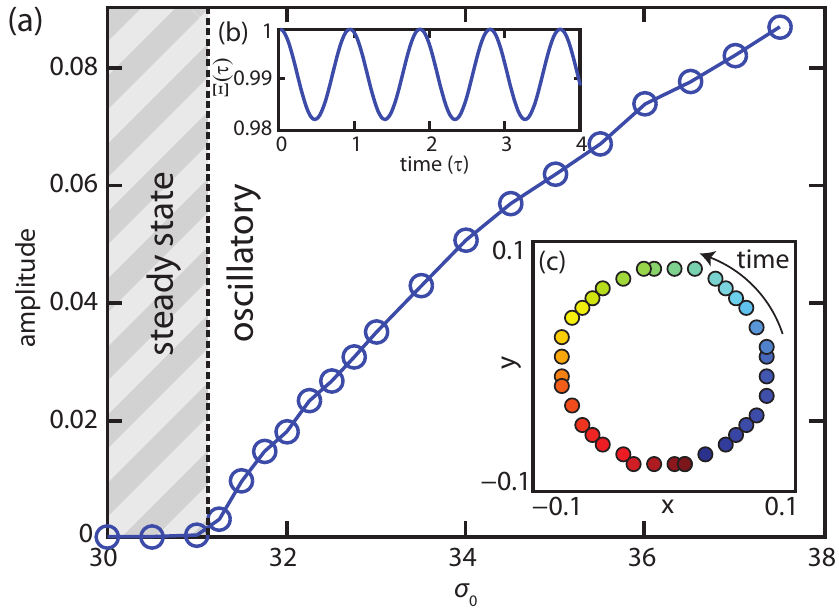}
\caption{(color online). Secondary bifurcation to oscillatory dynamics.  (a) Amplitude of oscillation of the 
velocity correlation function $\Xi(\tau)$ vs. $\sigma_0$, with $d^{(r)} = d^{(s)} = 0.025$, 
showing a bifurcation from steady defect separation to oscillatory behavior at a critical value of 
$\sigma_0$. (b) $\Xi(\tau)$ for $\sigma_0 = 32$ showing periodic oscillatory behavior. 
(c) Position of flow circulation center over time for $\sigma_0 = 32.75$ during one oscillation period.}
\label{fig:oscillation_amplitude}
\end{figure}

As $\sigma_0$ is increased beyond a new critical value, the steady state is unstable and the 
system bifurcates into a regime of 
\emph{periodic oscillation}, where the time symmetry is broken. 
The $+1/2$ defect pair (Fig.~\ref{fig:behaviour_spectrum}c) execute periodic `orbits' around each other, 
with the flow circulation center offset from the origin (Fig.~\ref{fig:behaviour_spectrum}f), following a 
quasi-circular trajectory (Fig.~\ref{fig:oscillation_amplitude}c). These states can be analyzed by 
examining the correlation function \cite{Schaller2010}
\begin{align*}
 \Xi(\tau) \equiv \left\langle \frac{\langle \bv(\bx,t) \cdot \bv(\bx,t-\tau) \rangle_\bx}{\langle \bv(\bx,t) 
\cdot \bv(\bx,t) \rangle_\bx} \right\rangle_t
\end{align*}
where the temporal average is taken 
over late times when the oscillatory state is fully established.  Extracting the amplitude $A$ of 
oscillation of $\Xi$ (Fig.~\ref{fig:oscillation_amplitude}b) we numerically determine a bifurcation 
diagram as a function of $\sigma_0$ as in Fig.~\ref{fig:oscillation_amplitude}a.  There is a 
clear threshold for the onset of periodic oscillations.  A similar oscillatory bifurcation has been observed 
by Giomi {\it et al.} \cite{Giomi} for a dense active nematic in a channel geometry, suggesting 
that such behavior may be a fundamental property of active nematics, though the exact form 
taken will be heavily dependent on geometry and topology. Were this system an annulus, rather than 
a disc, behavior more closely resembling the `back and forth' oscillations of \cite{Giomi} could be 
conjectured as a regime beyond the spontaneous flow of \cite{Furthauer2012}.

Motivated by principles of cytoplasmic streaming, we have constructed a clean, simple model for a 
dilute suspension of extensile force-generating filaments in total geometric confinement, and 
have demonstrated that inclusion of elementary hydrodynamics is entirely sufficient to yield 
spontaneous self-organization, in spite of the absence of more complex local interaction terms. 
In an experimental realization, the prediction of a critical activity for transition from quiescence to 
circulation can be tested by varying the viscosity or motor activity, perhaps through temperature
or ATP concentration.  Modern realizations of the Yotsuyanagi's experiment could provide a
wealth of information on this type of bifurcation. 

\begin{acknowledgments}
We thank S. Ganguly, E.J. Hinch, A. Honerkamp-Smith, P. Khuc Trong and H. Wioland for discussions. 
This work was supported by the EPSRC and European Research Council Advanced Investigator 
Grant 247333.
\end{acknowledgments}

\end{document}